\documentstyle[12pt]{article}
\input epsf
\epsfverbosetrue
\newlength{\dinwidth}
\newlength{\dinmargin}
\begin{document}
\def\deta{$\Delta\eta$}
\def\gapf{$f(\Delta\eta$}
\vspace*{2.0 cm}
\begin{center}
{\Large {\bf Rapidity Gaps between Jets in Photoproduction at HERA}}
\end{center}

\vspace{2cm}

\begin{center}
\begin{large}
 ZEUS Collaboration\\
\end{large}
\end{center}

\vspace{5cm}

\begin{abstract}

Photoproduction events which have two or more jets have been studied in the
$W_{\gamma p}$ range 135~GeV $< W_{\gamma p} <$ 280~GeV with the ZEUS
detector at HERA.  A class of events is observed with little hadronic
activity between the jets. The jets are separated by pseudorapidity intervals
($\Delta\eta$) of up to four units and have transverse energies greater than
6~GeV.  A gap is defined as the absence between the jets of particles with
transverse energy greater than 300~MeV.  The fraction of events containing a
gap is measured as a function of \deta.  It decreases exponentially as
expected for processes in which colour is exchanged between the jets, up to a
value of $\Delta\eta \sim 3$, then reaches a constant value of about 0.1. The
excess above the exponential fall-off can be interpreted as evidence for hard
diffractive scattering via a strongly interacting colour singlet object.

\end{abstract}

\vspace{-20cm}
\begin{flushleft}
\tt DESY 95-194 \\
October 1995 \\
\end{flushleft}

\setcounter{page}{0}
\thispagestyle{empty}
\newpage

%   23/10/95            MEMBER NAME  AUTH29   (TEX)      M  TEX
%   07/10/95 510171339  MEMBER NAME  AUTH029  (ZEUS)     M  TEX
%
\def\3{\ss}
\textwidth 15.5cm
\parindent 0cm
\footnotesize
\renewcommand{\thepage}{\Roman{page}}
\begin{center}
\begin{large}
The ZEUS Collaboration
\end{large}
\end{center}
M.~Derrick, D.~Krakauer, S.~Magill, D.~Mikunas, B.~Musgrave,
J.~Repond, R.~Stanek, R.L.~Talaga, H.~Zhang \\
{\it Argonne National Laboratory, Argonne, IL, USA}~$^{p}$\\[6pt]
G.~Bari, M.~Basile,
L.~Bellagamba, D.~Boscherini, A.~Bruni, G.~Bruni, P.~Bruni, G.~Cara
Romeo, G.~Castellini$^{1}$, M.~Chiarini,
L.~Cifarelli$^{2}$, F.~Cindolo, A.~Contin, M.~Corradi,
I.~Gialas$^{3}$,
P.~Giusti, G.~Iacobucci, G.~Laurenti, G.~Levi, A.~Margotti,
T.~Massam, R.~Nania, C.~Nemoz, F.~Palmonari, \\
A.~Polini, G.~Sartorelli, R.~Timellini, Y.~Zamora Garcia$^{4}$,
A.~Zichichi \\
{\it University and INFN Bologna, Bologna, Italy}~$^{f}$
% ...
\\[6pt]
\clearpage
\hspace*{1mm}
$^{ 1}$ also at IROE Florence, Italy  \\
\hspace*{1mm}
$^{ 2}$ now at Univ. of Salerno and INFN Napoli, Italy  \\
\hspace*{1mm}
$^{ 3}$ supported by EU HCM contract ERB-CHRX-CT93-0376 \\
\hspace*{1mm}
$^{ 4}$ supported by Worldlab, Lausanne, Switzerland  \\
\hspace*{1mm}
% ...
$^{21}$ now with OPAL Collaboration, Faculty of Physics at Univ. of
        Freiburg \\
% ...

\begin{tabular}{lp{15cm}}
$^{a}$ & supported by the Natural Sciences and Engineering Research
         Council of Canada (NSERC) \\
$^{b}$ & supported by the FCAR of Qu\'ebec, Canada\\
$^{c}$ & supported by the German Federal Ministry for Education and
         Science, Research and Technology (BMBF), under contract
         numbers 056BN19I, 056FR19P, 056HH19I, 056HH29I, 056SI79I\\
$^{d}$ & supported by the MINERVA Gesellschaft f\"ur Forschung GmbH,
         and by the Israel Academy of Science \\
$^{e}$ & supported by the German Israeli Foundation, and
         by the Israel Academy of Science \\
$^{f}$ & supported by the Italian National Institute for Nuclear Physics
         (INFN) \\
$^{g}$ & supported by the Japanese Ministry of Education, Science and
         Culture (the Monbusho)
         and its grants for Scientific Research\\
$^{h}$ & supported by the Korean Ministry of Education and Korea Science
         and Engineering Foundation \\
$^{i}$ & supported by the Netherlands Foundation for Research on Matter
         (FOM)\\
$^{j}$ & supported by the Polish State Committee for Scientific
         Research, grants No.~115/E-343/SPUB/P03/109/95, 2P03B 244
         08p02, p03, p04 and p05, and the Foundation for Polish-German
         Collaboration (proj. No. 506/92) \\
$^{k}$ & supported by the Polish State Committee for Scientific
         Research (grant No. 2 P03B 083 08) \\
$^{l}$ & partially supported by the German Federal Ministry for
         Education and Science, Research and Technology (BMBF) \\
$^{m}$ & supported by the German Federal Ministry for Education and
         Science, Research and Technology (BMBF), and the Fund of
         Fundamental Research of Russian Ministry of Science and
         Education and by INTAS-Grant No. 93-63 \\
$^{n}$ & supported by the Spanish Ministry of Education and Science
         through funds provided by CICYT \\
$^{o}$ & supported by the Particle Physics and Astronomy Research
         Council \\
$^{p}$ & supported by the US Department of Energy \\
$^{q}$ & supported by the US National Science Foundation
\end{tabular}

\newpage
\pagenumbering{arabic}
\setcounter{page}{1}
\normalsize

\section{Introduction}

In high energy hadronic collisions, the dominant mechanism for jet production
is described by a hard scatter between partons in the incoming hadrons via a
quark or gluon propagator. This propagator carries colour charge. Since
colour confinement requires that the final state contain only colour singlet
objects, the exchange of colour quantum numbers in the hard process means
that a jet at some later stage generally exchanges colour with another jet or
beam remnant widely separated from it in rapidity. Such jets are said to be
``colour connected'' and this leads to the production of particles throughout
the rapidity region between the jets.  However, if the hard scattering were
mediated by the exchange of a colour singlet propagator in the $t$-channel,
each jet would be colour connected only to the beam remnant closest in
rapidity and the rapidity region between the jets would contain few
final-state particles~\cite{dokshitzer}. The colour singlet propagator could
be an electroweak gauge boson or a strongly interacting object, and the soft
gluon emission pattern produced in each case is similar~\cite{zeppenfeld}.
However the rates could be very different. In order to determine the rate of
colour singlet exchange processes it has been proposed~\cite{bjorken} to
study the multiplicity distribution in pseudorapidity\footnote{$\eta = -
$ln$(\tan\frac{\vartheta}{2}$) where $\vartheta$ is the polar angle with
respect to the $z$ axis, which in the ZEUS coordinate system is defined to be
the proton direction.} ($\eta$) and azimuth ($\varphi$) of the final state
particles in dijet events, and to count events with an absence of particles
(i.e. with a rapidity gap) between the two jets.

D0~\cite{D0} and CDF~\cite{CDF} have reported the results of searches in
$p\bar{p}$ collisions at $\sqrt{s} = 1.8$~TeV for dijet events containing a
rapidity gap between the two highest transverse energy ($E_T^{jet}$) jets.
Both collaborations see an excess of gap events over the expectations from
colour exchange processes. D0 report an excess of $0.0107 \pm 0.0010(stat.)
^{+0.0025} _{-0.0013}(sys.)$, whereas CDF measure the fraction to be $0.0086
\pm 0.0012$. We report here the results of a similar search in $\gamma p$
interactions obtained from $e^{+}p$ collisions at HERA.

In leading order, two processes are responsible for jet production in $\gamma
p$ interactions at HERA.  In the first case, the direct contribution, the
photon interacts directly with a parton in the proton. In the second case,
the resolved contribution, the photon first fluctuates into a hadronic state
which acts as a source of partons which then scatter off partons in the
proton. Fig.~1(a) shows schematically an example of colour singlet exchange
in resolved photoproduction in which a parton in the photon scatters from a
parton in the proton, via $t$-channel exchange of a colour singlet object. An
example of the more common colour non-singlet exchange mechanism is shown in
Fig.~1(b). For high $E_T^{jet}$ dijet production, the magnitude of the square
of the four-momentum ($|t|$) transferred by the colour singlet object is
large. Thus it is possible to calculate in perturbative QCD the cross section
for the exchange of a strongly interacting colour singlet
object~\cite{bjorken,muellertang,delduca,hjl}. For instance, the ratio of the
two-gluon colour singlet exchange cross section to the single gluon exchange
cross section has been estimated to be about 0.1~\cite{bjorken}. Studies of
rapidity gaps at high $|t|$ (``hard diffractive scattering'') are
complementary to studies of diffractive hard scattering where the rapidity
gap is between a colourless beam remnant, produced with low four-momentum
transfer with respect to one of the beam particles, and hadronic activity in
the central detector~\cite{forwardg}.

The event morphology for the process of Fig.~1(a) is illustrated in
Fig.~1(c). There are two jets in the final state, shown as circles in
($\eta,\varphi$) space. Here \deta~ is defined as the distance in $\eta$
between the centres of the two jet cones. For the colour singlet exchange
process of Fig.~1(a), radiation into the region (labelled ``gap'') between
the jet cones is suppressed, giving rise to the rapidity gap signature.
Multiplicity fluctuations in colour non-singlet exchange events can also
produce gaps between jets. In order to disentangle the different mechanisms
for gap production it is useful to introduce the concept of the
`gap-fraction'.

The gap-fraction, $f(\Delta\eta)$, is defined as the ratio of the number of
dijet events at this \deta\ which have a rapidity gap between the jets to the
total number of dijet events at this \deta. For colour non-singlet exchange,
the gap-fraction is expectated to fall exponentially with increasing \deta.
This exponential behaviour can be taken as a definition of non-diffractive
processes~\cite{bjorken}. The expectation follows from the assumption that
the probability density for radiation of a particle is constant across the
rapidity interval between the jets and it is consistent with the results of
analytic QCD calculations~\cite{delduca}, and with Monte Carlo simulation
(see subsequent sections). For colour singlet exchange, the gap-fraction is
not expected to depend strongly upon $\Delta\eta$~\cite{bjorken,delduca}.
Therefore, at sufficiently large \deta, such a colour singlet contribution
will dominate the behaviour of the gap-fraction.  The situation is
illustrated in Fig.~1 (d), where the colour non-singlet contribution is shown
as an exponential fall-off, and the colour singlet contribution is shown as
independent of \deta.

In this paper the gap fraction is studied for a sample of photoproduction
events with two jets of $E_T^{jet} > 6$~GeV. The events are obtained from an
integrated luminosity of 2.6~pb$^{-1}$ of $e^{+} p$ collisions measured by
the ZEUS detector and have $\gamma p$ centre-of-mass energies in the range
135~GeV~$< W_{\gamma p} <$~280~GeV. Dijet cross sections are measured as a
function of \deta\ for events with a gap and for events with no gap
requirement.

\section{Experimental Setup}

Details of the ZEUS detector have been described elsewhere~\cite{ZEUS}.  The
primary components used in this analysis are the central calorimeter and the
central tracking detectors.  The uranium-scintillator calorimeter~\cite{CAL}
covers about 99.7\% of the total solid angle and is subdivided into
electromagnetic and hadronic sections with respective typical cell sizes, of
$5 \times 20$ cm$^2$ ($10 \times 20$ cm$^2$ in the rear calorimeter, i.e. the
positron direction) and $20 \times 20$ cm$^2$. The central tracking system
consists of a vertex detector~\cite{VXD} and a central tracking
chamber~\cite{CTD} enclosed in a 1.43~T solenoidal magnetic field.

A photon lead-scintillator calorimeter is used to measure the luminosity via
the positron-proton Brems\-strahlung process.  This calorimeter is installed
inside the HERA tunnel and subtends a small angle in the positron beam
direction from the interaction vertex~\cite{LUMI}.  Low angle scattered
positrons are detected in a similar lead-scintillator calorimeter.

In 1994 HERA provided 820~GeV protons and 27.5~GeV positrons colliding in 153
bunches. Additional unpaired positron and proton bunches circulated to allow
monitoring of background from beam-gas interactions.

\section{Data Selection}

The ZEUS data acquisition uses a three level trigger system. At the first
level events are selected which were triggered on a coincidence of a regional
or transverse energy sum in the calorimeter with a track coming from the
interaction region measured in the central tracking chamber.  At the second
level a cut was made on the total transverse energy, and cuts on calorimeter
energies and timing were used to suppress events caused by interactions
between the proton beam and residual gas in the beam pipe~\cite{F2}. At the
third level, tracking cuts were made to reject events arising from proton
beam-gas interactions and cosmic ray events. Also at the third level, jets
were found from the calorimeter cell energies and positions using a fast cone
algorithm and events were required to have at least two jets.

Charged current events are rejected by a cut on the missing transverse
momentum measured in the calorimeter. Events with a scattered positron
candidate in the calorimeter are rejected. This restricts the range of the
photon virtuality to $P^{2} < 4$~GeV$^{2}$, and results in a median $P^{2}$
of $\sim 10^{-3}$~GeV$^2$.  A cut of $0.15 \leq y < 0.7$ is applied on the
fraction of the positron's momentum which is carried by the photon, where $y$
is reconstructed using the Jacquet-Blondel method~\cite{YJB}. This cut
restricts the $\gamma p$ centre-of-mass energies to lie in the range
135~GeV~$< W_{\gamma p} <$~280~GeV.

To select the final jet sample, a cone algorithm~\cite{snow} is applied to
the calorimeter cells.  Cells within a radius $R = \sqrt{\delta\eta^2_{cell}
+ \delta\varphi^2_{cell}}$ of 1.0 from the jet centre are included in the jet
where $\delta\eta^{cell}$ amd $\delta\phi^{cell}$ represent respectively the
difference in pseudorapidity and azimuthal angle (in radians) between the
centre of the cell and the jet axis. Events are then required to have at
least two jets found in the uranium calorimeter with $E_T^{jet} > 5$~GeV and
$\eta^{jet} < 2.5$.  In addition the two highest transverse energy
jets\footnote{In \cite{delduca} the jets are ordered in pseudorapidity rather
than transverse energy and the two jets at lowest and highest pseudorapidity
are used in the calculation.  When the uncorrected gap-fraction is made with
this selection, it is about 0.01 lower.} were required to have $\Delta\eta >
2$ (i.e. cones not overlapping in $\eta$) and boost $|(\eta_1 + \eta_2)|/2 =
|\bar{\eta}| < 0.75$.  These conditions constrain the jets to lie within the
kinematic region where the detector and event simulations are best
understood.

To identify gap events, the particle multiplicity is determined by grouping
calorimeter cells into ``islands''.  This is done by assigning to every cell
a pointer to its highest energy neighbour.  A cell which has no highest
energy neighbour is a local maximum.  An island is formed for each local
maximum which includes all of the cells that point to it. The events with
{\it no} islands of transverse energy $E_T^{island} > 250$~MeV and $\eta$
between the edges of the jet cones (as defined by the cone radius $R$) are
called gap events.

A total of $8393$ dijet events were selected, of which $3186$ are gap events.
The non-$e^{+}p$ collision background was estimated using the number of
events associated with unpaired bunch crossings. The beam gas background was
found to be less than 0.1\%. The cosmic ray contamination is estimated to be
about 0.1\%. For those events in which the low angle scattered positron is
detected in lead-scintillator calorimeter, $P^2 < 0.02$~GeV$^2$. The fraction
of these events is around 20\%, in agreement with the Monte Carlo
expectation. The 43 gap events which have $\Delta\eta > 3.5$ were also
scanned visually to search for contamination from events where the energy
deposits of the scattered positron or a prompt photon might mimic a jet. No
such events were found.

\section{Results}

In section~4.1 we present results obtained from ZEUS data which are not
corrected for detector effects, by comparing the data to Monte Carlo
generated events which have been passed through a detailed simulation of the
ZEUS detector and selection criteria. The PYTHIA~\cite{pythia} Monte Carlo
program was used with the minimum $p_T$ of the hard scatter set to 2.5~GeV.
The GRV~\cite{GRV} parton distributions were used for the photon and the
MRSA~\cite{MRSA} parton distributions were used for the proton.  Two Monte
Carlo event samples were generated. For the first sample (``PYTHIA
non-singlet''), resolved and direct photon interactions were generated
separately and combined according to the cross sections determined by PYTHIA.
No electroweak exchange (quark quark scattering via $\gamma/Z^0$ or $W^{\pm}$
exchange) events were included. For the second sample (``PYTHIA mixed''),
10\% of electroweak exchange events were included. This fraction is two
orders of magnitude higher than the level actually expected from the cross
section for these events and is chosen in order to mimic the effect of
strongly interacting colour singlet exchange processes which are not included
in PYTHIA.

In section~4.2 we present the ZEUS data after corrections for all detector
acceptance and resolution effects.  These hadron-level measurements are then
compared to model predictions, and to the expectation of an exponential
suppression of gap production for non-diffractive processes.

\subsection{Uncorrected Results}

The energy flow $1 / N dE_T^{cell}/d\delta\eta^{cell}$ with respect to the
jet axis for cells within one radian in $\varphi$ of the jet axis is shown in
Fig.~2(a) for the two highest transverse energy jets of each event. PYTHIA
mixed events are shown as the solid line. Here and throughout Fig.~2 the data
are shown as black dots and the errors shown are statistical only. This jet
profile shows highly collimated jets in the data and a pedestal of less than
1~GeV of transverse energy per unit pseudorapidity outside the jet cone
radius of 1.0. The pedestal transverse energy is higher toward the proton
direction. The superposition of profiles of one jet at high $\eta^{jet}$ and
one at low $\eta^{jet}$ leads to the bump at $\delta\eta^{cell} \sim 1.5$,
due to the forward edge of the calorimeter.  The profiles for the PYTHIA
non-singlet sample are not shown, as they are similar to those of the mixed
sample. The PYTHIA events generally describe the data well, although they are
slightly more collimated and underestimate the forward jet pedestal. This
small discrepancy may be related to secondary interactions between the photon
remnant and the proton remnant, which are not simulated in these PYTHIA
samples. Including any kind of multiple interactions in the simulation
increases the energy flow and particle multiplicity~\cite{MI} and thus can
only decrease the number of gaps predicted by the Monte Carlo program.

The distribution of the total number of events (without any demand on the
presence or absence of a gap) as a function of \deta\ is shown in Fig.~2(b).
It decreases with increasing $\Delta\eta$, extending out to $\Delta\eta \sim
4$.  The PYTHIA distributions are normalized to the number of events in the
data. Both PYTHIA samples provide an adequate description of this
distribution although the total number of events seen at large \deta~ is
slightly underestimated.

The distribution of the gap events as a function of \deta~ is plotted in
Fig.~2(c) where the normalisation for the PYTHIA distribution is the same as
in Fig.~2(b). The number of events in the data exhibiting a gap falls steeply
with $\Delta\eta$. However the expectation from the PYTHIA non-singlet sample
falls more steeply than the data, significantly underestimating the number of
gap events at large $\Delta\eta$.  The PYTHIA sample with a mixture of 10\%
of electroweak boson exchange can account for the number of gap events in the
data at large $\Delta\eta$. However this sample significantly overestimates
the number of gap events at low $\Delta\eta$. As mentioned previously,
including secondary interactions in the simulation could reduce the predicted
number of gap events and possibly account for this discrepancy.

By taking the ratio of Fig.~2(c) to Fig.~2(b), the gap-fraction shown in
Fig.~2(d) is obtained. The gap-fraction falls exponentially out to
$\Delta\eta \sim 3.2$. Thereafter it levels off at a value of roughly 0.08.
The PYTHIA non-singlet sample fails to describe the flat region in the data,
falling approximately exponentially over the whole measured range of \deta.
This sample also overestimates the fraction of gap events at low
$\Delta\eta$.  The PYTHIA mixed sample can describe the flat region of the
data but again overestimates the gap-fraction at low $\Delta\eta$.  The
gap-fraction for the electroweak exchange events alone exceeds 0.4 over the
full $\Delta\eta$ range (not shown).

The uncorrected data exhibit a flat region at large \deta\ consistent with a
colour singlet contribution of around 10\%.  Detector effects are expected to
largely cancel in the gap-fraction. In the next section we find that this is
indeed the case and provide quantitative estimates of both the discrepancy
between PYTHIA and the data and of the significance of the deviation of the
measured gap-fraction from an exponential fall.

\subsection{Corrected Results}

In order to investigate whether the observed flat region in the gap-fraction
might be a detector effect, the PYTHIA mixed sample has been used to correct
the data for all detector effects, including acceptance, smearing and the
shift in the measurement of energies. Cross sections are determined and the
gap-fraction is measured in four bins of \deta~ in the range $2 \leq
\Delta\eta < 4$.

The cross section $d\sigma / d\Delta\eta$ is measured for dijet
photoproduction, $ep \rightarrow e\gamma p \rightarrow e X$, where $X$
contains at least two jets of final state particles. The cross section is
measured in the range $0.2 < y < 0.85$ for photon virtualities $P^2 <
4$~GeV$^2$.  The two jets are defined by a cone algorithm with a cone radius
of 1.0 in ($\eta, \varphi$) and satisfy $E_T^{jet} > 6$~GeV, $\eta^{jet} <
2.5$. The two jets of highest $E_T^{jet}$ satisfy $\Delta\eta > 2$ and
$|\bar{\eta}| < 0.75$.  The rear $\eta^{jet}$ distribution falls to zero at
$\eta^{jet} \sim -2$, well within the rear detector acceptance.  Therefore no
explicit rear pseudorapidity cut is made. The gap cross section,
$d\sigma_{gap} / d\Delta\eta$, is measured, in the same kinematic range, for
events with no final state particles with transverse energy $E_T^{particle} >
300$~MeV between the jet cones. The corrected gap-fraction $f(\Delta\eta)$ is
then obtained from the ratio of $d\sigma_{gap} / d\Delta\eta$ to $d\sigma /
d\Delta\eta$.

The efficiency of the data selection described in section~3 for finding
events in this kinematic region was estimated using the Monte Carlo samples.
The combined efficiency of the online triggers is at least 80\% in every bin
of $\Delta\eta$. The efficiency of the offline selection is about 50\%
leading to a combined efficiency of the online and offline selection criteria
of about 40\%. The low efficiency of the offline selection is due to the
finite detector resolution of the jet energy and angular variables, and the
steeply falling $E_T^{jet}$ spectrum. The shifts and resolutions of these
variables are consistent with those obtained in extensive studies of the 1993
dijet sample~\cite{dijets}.  The $E_T^{particle}$ resolution is 27\% with a
shift of -14\%.  The $\eta^{particle}$ resolution is 0.01 with negligible
shift.

The final correction factors for the inclusive cross section are smoothly
varying between 1.6 in the lowest \deta~ bin and 1.4 in the highest \deta~
bin. The correction factors for the gap cross section are between 1.5 and
1.8. The ratios of these correction factors form effective correction factors
for the gap-fraction which are all within 27\% of unity.

The systematic uncertainties have been estimated by varying the cuts made on
the reconstructed quantities. The island algorithm for counting particles was
replaced by an algorithm which clusters cells based on proximity in ($\eta,
\varphi$) space. Also the results of the bin-by-bin correction were checked
using an unfolding procedure~\cite{giulio}. The correction was also performed
by using the PYTHIA non-singlet sample and by leaving out the leading order
direct contribution.  The uncertainty due to the parton distribution was
included. The uncertainty due to a 3.3\% systematic error in the luminosity
measurement was included. Finally the systematic uncertainty arising from a
5\% uncertainty in the mean energies measured by the calorimeter was
estimated.  This represents the largest uncertainty in the two cross sections
but cancels in the gap-fraction.  The largest systematic uncertainty which
remains in the gap-fraction comes from the variation of the $E_T^{island}$
cut from 200 to 300~MeV. The combined effect of these uncertainties is
included in the outer error bars in Fig.~3.

The inclusive and gap cross sections and the corrected gap-fraction as a
function of \deta\ are presented in Figs.~3(a) to (c) respectively (black
dots) and compared with the expectations from the PYTHIA non-singlet exchange
cross sections (open circles). For the data, the inner error bars show the
statistical errors and the outer error bars display the systematic
uncertainties, added in quadrature. The cross section points are plotted at
the centres of the bins.  The gap-fraction points are plotted at the mean
$\Delta\eta$ values of the inclusive cross section. Numerical values for the
inclusive cross section, the gap cross section and the corrected gap-fraction
are provided in Tables 1, 2 and 3, respectively.

The inclusive cross section is around 5~nb per unit \deta\ at \deta~$=2$,
falling to about 0.5~nb for $\Delta\eta > 3.5$. The gap cross section is
around 3~nb per unit \deta\ at \deta~$=2$ and falls to about 0.06~nb for
\deta~ between 3.5 and 4. The overall normalization of PYTHIA agrees with the
data within the errors. PYTHIA also describes the shape of the inclusive
cross section. However it fails to describe the gap cross section, falling
too steeply with \deta~ and disagreeing significantly in the last bin.

The corrected gap-fraction falls exponentially in the first three bins but
the height of the fourth bin is consistent with the height of the third.  The
height of the fourth bin is $0.11 \pm 0.02(stat.) ^{+0.01} _{-0.02}(sys.)$,
which is also consistent with the flat region at large \deta\ seen in the
uncorrected gap fraction and inconsistent with the expectation from PYTHIA.

\section{Discussion}

Two methods have been used to estimate the significance of the excess of the
gap-fraction over the expectation from multiplicity fluctuations in
non-singlet exchange.

The first method is to take the difference between the data and the PYTHIA
non-singlet gap-fractions, shown in Fig.~3(c).  An excess of $0.07 \pm 0.03$
is obtained, based entirely on the last bin. However this is a
model-dependent estimate. For instance replacing the Lund symmetric
fragmentation function by the Field-Feynman fragmentation function yields a
lower predicted gap-fraction and a larger excess.  Introducing multiple
interactions into PYTHIA also lowers the fraction of gap events expected. On
the other hand, lowering $\sigma_{p_T}$ (which controls the hadron momentum
distribution transverse to the parent parton) in the Monte Carlo simulation
from the default value of 0.36~GeV to 0.25~GeV produces a gap-fraction that
is very like the data.  It has a height in the fourth bin of $0.07 \pm 0.02$
and therefore if one believes this model, there is no significant excess.
However this option yields jet-profiles which are narrower than the default
PYTHIA profiles, which are already slightly narrower than the data.

The second way to estimate the excess of the gap-fraction over that expected
from purely non-singlet exchange does not rely on comparisons to Monte Carlo
predictions. In Fig.~3(d) the gap-fraction is shown again and compared with
the result of a two parameter ($\alpha$,$\beta$) $\chi^2$-fit to the
expression
\begin{math}
  f^{fit}(\alpha,\beta;\Delta\eta) = C(\alpha,\beta) e^{\alpha \Delta\eta} +
\beta
\end{math}
where $C(\alpha,\beta)$ is the normalization coefficient constraining
$f^{fit}(\alpha,\beta;\Delta\eta)$ to 1.0 at $\Delta\eta = 2$. The result of
this fit is shown as the solid curve in Fig~3(d), and the exponential (dotted
line) and constant (dashed line) terms are also shown. The quality of this
fit, as indicated by the $\chi^2$ value of 1.2 for the two degrees of freedom
is superior to that of a fit to an exponential alone which yields $\chi^2 =
9$.  The fit parameters are $\alpha = -2.7 \pm 0.3(stat.) \pm 0.1(sys.)$ and
$\beta = 0.07 \pm 0.02(stat.) ^{+0.01} _{-0.02}(sys.)$. The parameter $\beta$
gives an estimate of the gap-fraction for colour singlet processes. This
method uses the full information of the four measured data points and is not
dependent on the details of the Monte Carlo fragmentation model. However, the
assumption that the colour singlet gap-fraction is constant with \deta\ is
only one of many possibilities.

Both the comparison with the default PYTHIA non-singlet prediction and the
fit to an exponential form give an excess of about 0.07 in the gap-fraction
over the expectation from colour non-singlet exchange.

The excess in the gap-fraction over the expectation from non-singlet exchange
may be interpreted as evidence for the exchange of a colour singlet object.
In fact the fraction of events due to colour singlet exchange,
$\hat{f}(\Delta\eta)$, may be even higher than the measured excess.  As
previously mentioned, secondary interactions of the photon and proton remnant
jets could fill in the gap.  A survival probability, ${\cal P}$, has been
defined~\cite{bjorken} which represents the probability that a secondary
interaction does not occur.  Then $f(\Delta\eta) = \hat{f}(\Delta\eta) \cdot
{\cal P}$.  Estimates of the survival probability for $p\bar{p}$ collisions
at the Tevatron range from about 5\% to 30\%~\cite{bjorken,GLM,fletcher}.
The survival probability at HERA could be higher due to the lower
centre-of-mass energy, the fact that one remnant jet comes from a photon
rather than a proton and the fact that the mean fraction of the photon energy
participating in the jet production in these events is high\footnote{The
average fraction of the photon energy participating in the production of the
two jets~\cite{dijets} is 0.7 for these events. Nevertheless, according to
the PYTHIA simulation the dominant contribution in this kinematic regime is
from leading order resolved events.}. Therefore the ZEUS result of $\sim
0.07$ and the D0 and CDF results of $0.0107 \pm 0.0010(stat.) ^{+0.0025}
_{-0.0013}(sys.)$ and $0.0086 \pm 0.0012$ could arise from the same
underlying process.

The magnitude of the squared four-momentum transfer across the rapidity gap
as calculated from the jets is large ($|t| \ge (E_T^{jet})^2$).  Thus the
colour singlet exchange is unambiguously ``hard''.

The PYTHIA generator predicts that the ratio of the electroweak
($\sigma^{EW}$) to QCD ($\sigma^{QCD}$) exchange cross sections in this
kinematic range is $\sigma^{EW} / \sigma^{QCD} < 7 \cdot 10^{-4}$ (compatible
with the estimation $(\alpha / \alpha_{s})^2$).  Therefore quark quark
scattering via $\gamma/Z^{0}$ and $W^{\pm}$ exchange cannot explain the
height of the flat region in the gap-fraction. On the other hand, using the
simple two-gluon model for pomeron exchange gives $\hat{f}(\Delta\eta) \sim
0.1$~\cite{bjorken}. Thus pomeron exchange could account for the data.

In summary, dijet photoproduction events with $E_T^{jet} > 6$~GeV
contain an excess of events with a rapidity gap between the two
jets over the expectations of colour exchange processes.
This excess is observed as a flat region in the gap-fraction
at large rapidity separation (\deta = 3.7) at a level of
$0.11 \pm 0.02(stat.) ^{+0.01} _{-0.02}(sys.)$.
It can be interpreted as evidence of hard diffractive
scattering via a strongly interacting colour singlet object.

\section*{Acknowledgements}
We thank the DESY Directorate for their strong support and
encouragement and the HERA machine group for providing
colliding beams. We acknowledge the assistance of the DESY
computing and networking staff.  It is also a pleasure to thank
V. Del Duca for useful discussions.

%--------- REFERENCES -------------

%%%%%%%%%%%TABLES BEGIN HERE%%%%%%%%%%%%%%%%%
\clearpage
\begin{table}
\centering
\begin{tabular}{|r|c|c|c|}  \hline
$\Delta\eta$\rule[-.2cm]{0cm}{.7cm} & $d\sigma/d\Delta\eta$ & Statistical &
Systematic          \\
                             &         (nb)          & Uncertainty (nb)  &
Uncertainty (nb)          \\ \hline
    2.25    \rule[-.2cm]{0cm}{.7cm} & 4.93                  & 0.24        &
$^{+0.83} _{-0.68}$ \\
    2.75    \rule[-.2cm]{0cm}{.7cm} & 3.06                  & 0.15        &
$^{+0.54} _{-0.52}$ \\
    3.25    \rule[-.2cm]{0cm}{.7cm} & 1.67                  & 0.07        &
$^{+0.31} _{-0.19}$ \\
    3.75    \rule[-.2cm]{0cm}{.7cm} & 0.54                  & 0.03        &
$^{+0.08} _{-0.03}$ \\ \hline
\end{tabular}
\caption{$d\sigma/d\Delta\eta$ for $ep \rightarrow
e\gamma p \rightarrow e X$ in the kinematic range
$0.2 < y < 0.8$ and
$P^2 < 4$~GeV$^2$ and
where $X$ contains two or more jets of
$E_T^{jet} > 6$~GeV,
$\eta^{jet} < 2.5$,
$|\bar{\eta}| < 0.75$ and
$\Delta\eta > 2$.}
\end{table}

\begin{table}
\centering
\begin{tabular}{|c|c|c|c|}  \hline
$\Delta\eta$\rule[-.2cm]{0cm}{.7cm} & $d\sigma^{gap}/d\Delta\eta$ & Statistical
& Systematic          \\
                             &         (nb)                & Uncertainty (nb)
& Uncertainty (nb)          \\ \hline
    2.25    \rule[-.2cm]{0cm}{.7cm} & 2.85                        & 0.17
& $^{+0.45} _{-0.45}$ \\
    2.75    \rule[-.2cm]{0cm}{.7cm} & 0.66                        & 0.06
& $^{+0.11} _{-0.15}$ \\
    3.25    \rule[-.2cm]{0cm}{.7cm} & 0.16                        & 0.02
& $^{+0.03} _{-0.04}$ \\
    3.75    \rule[-.2cm]{0cm}{.7cm} & 0.06                        & 0.01
& $^{+0.01} _{-0.01}$ \\ \hline
\end{tabular}
\caption{$d\sigma^{gap}/d\Delta\eta$ for $ep \rightarrow
e\gamma p \rightarrow e X$ in the kinematic range
$0.2 < y < 0.8$ and
$P^2 < 4$~GeV$^2$ and
where $X$ contains two or more jets of
$E_T^{jet} > 6$~GeV,
$\eta^{jet} < 2.5$,
$|\bar{\eta}| < 0.75$ and
$\Delta\eta > 2$ with {\it no} final state particles
of $E_T^{particle} > 300$~MeV between the jets.}
\end{table}

\begin{table}
\centering
\begin{tabular}{|c|c|c|c|}  \hline
$\Delta\eta$\rule[-.2cm]{0cm}{.7cm} & $f(\Delta\eta)$       & Statistical &
Systematic          \\
                             &                       & Uncertainty       &
Uncertainty               \\ \hline
    2.23    \rule[-.2cm]{0cm}{.7cm} & 0.58                  & 0.04        &
$^{+0.04} _{-0.02}$ \\
    2.73    \rule[-.2cm]{0cm}{.7cm} & 0.22                  & 0.02        &
$^{+0.02} _{-0.02}$ \\
    3.22    \rule[-.2cm]{0cm}{.7cm} & 0.10                  & 0.01        &
$^{+0.01} _{-0.02}$ \\
    3.70    \rule[-.2cm]{0cm}{.7cm} & 0.11                  & 0.02        &
$^{+0.01} _{-0.02}$ \\ \hline
\end{tabular}
\caption{The gap-fraction, $f(\Delta\eta)$, for $ep \rightarrow
e\gamma p \rightarrow e X$ in the kinematic range
$0.2 < y < 0.8$ and
$P^2 < 4$~GeV$^2$ and
where $X$ contains two or more jets of
$E_T^{jet} > 6$~GeV,
$\eta^{jet} < 2.5$,
$|\bar{\eta}| < 0.75$ and
$\Delta\eta > 2$.}
\end{table}

%%%%%%%%%%FIGURES BEGIN HERE%%%%%%%%%%%%%
\clearpage
\begin{figure}
\centering
\leavevmode
\epsfbox{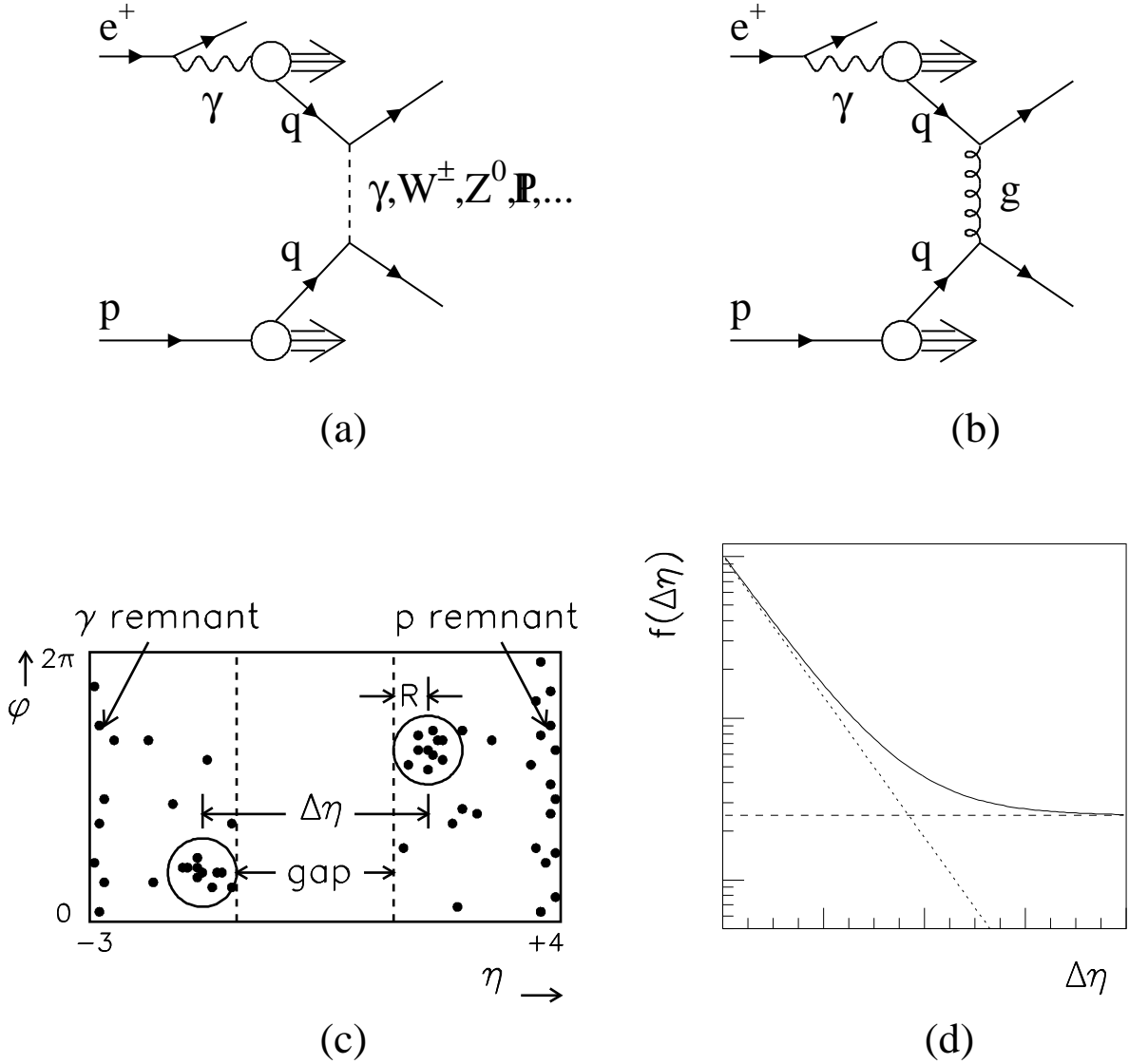}
\caption{
Resolved photoproduction via (a) colour singlet exchange and (b)
colour non-singlet exchange.
The rapidity gap event morphology is shown in (c) where black dots
represent final state hadrons and the boundary illustrates the limit
of the ZEUS acceptance.  Two jets of radius $R$ are shown, which are back
to back in azimuth and separated by a pseudorapidity interval
$\Delta\eta$.
An expectation for the behaviour of the gap fraction is shown
in (d)(solid line).
The non-singlet contribution is shown as the dotted line and
the colour singlet contribution as the dashed line.}
\end{figure}

\begin{figure}
\centering
\leavevmode
\epsfbox{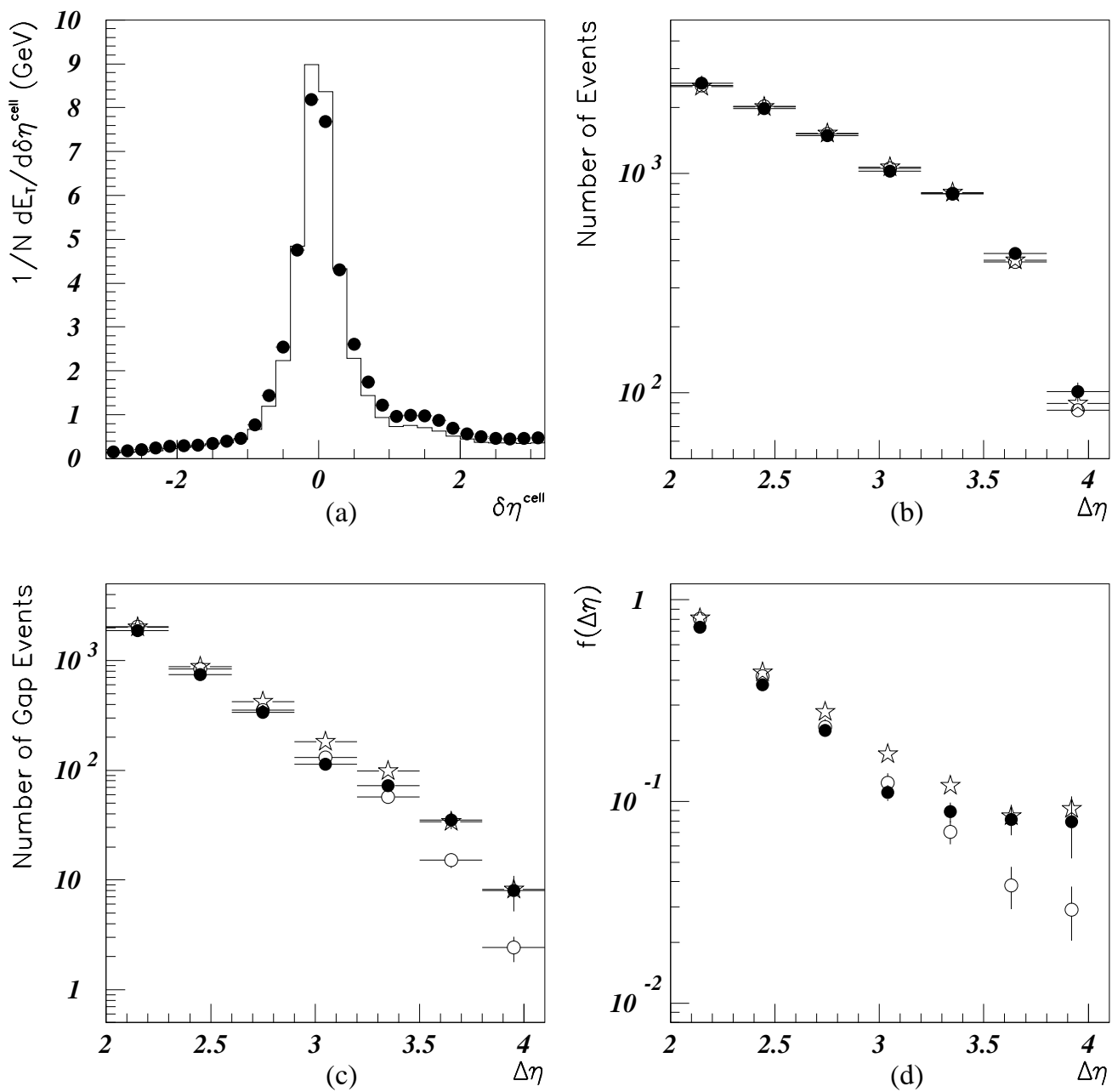}
\caption{Uncorrected data compared with the predictions from
PYTHIA events which have been passed through a detailed simulation of
the ZEUS detector and of the sample selection criteria.  The errors
shown are statistical only.
The transverse energy flow with respect to the jet axis is shown in (a)
where the data are shown as black dots and the PYTHIA non-singlet sample
is shown as a solid line.
In (b), (c) and (d) the data are again shown as black dots.  The
PYTHIA non-singlet sample is shown
as open circles and the PYTHIA mixed sample (which contains
10\% of colour singlet exchange events) is shown as stars.
The number of events versus $\Delta\eta$ is shown in (b).
The number of gap events versus $\Delta\eta$ is shown in (c) and
the gap-fraction is shown in (d). In (d) the points are
drawn at the mean \deta\ of the inclusive distribution in the
corresponding bin.
}
\end{figure}

\begin{figure}
\centering
\leavevmode
\epsfbox{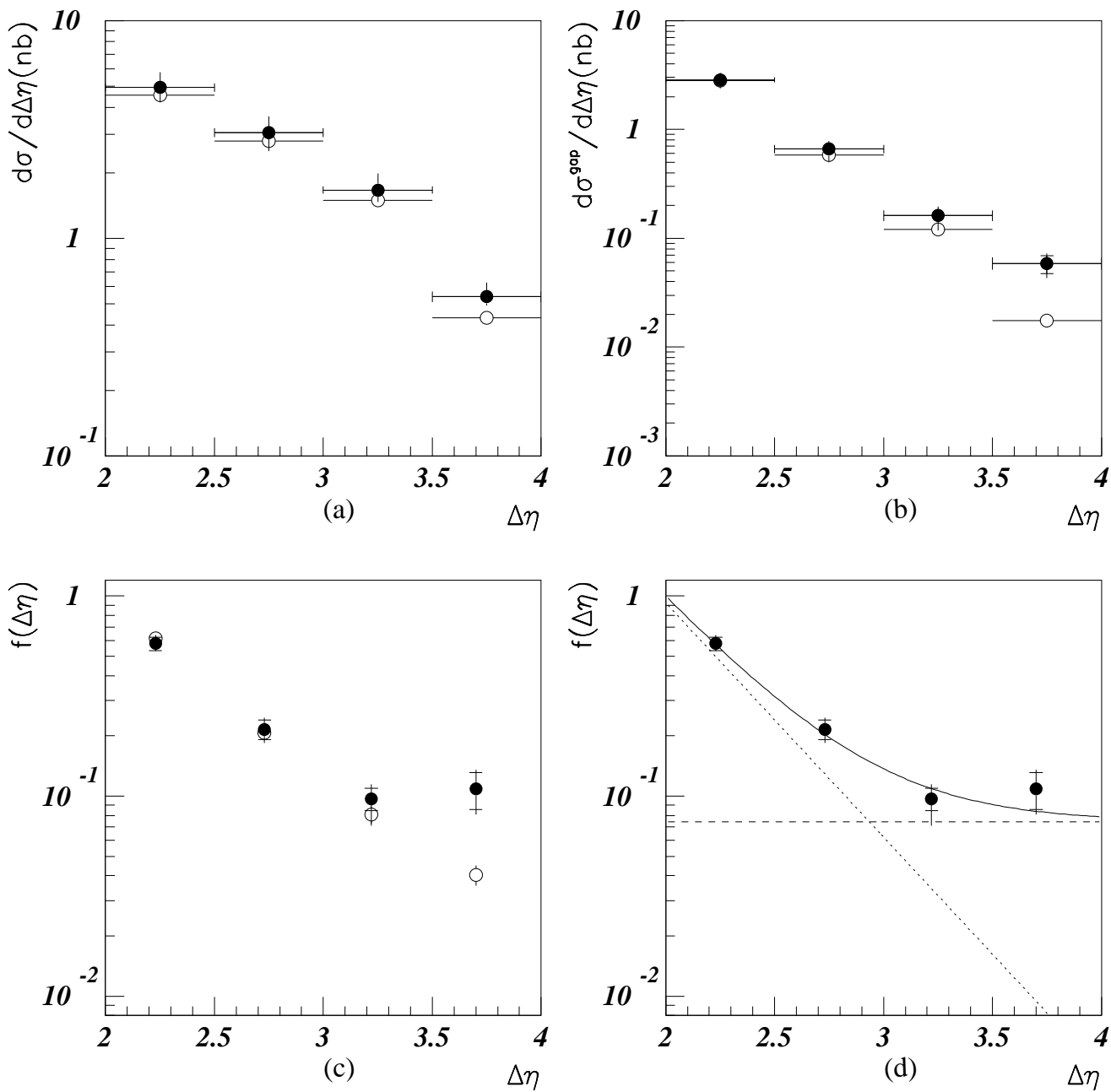}
\caption{ZEUS data (black circles) corrected for detector effects.
The inner error bars represent the statistical errors from the
data and Monte Carlo samples, and the outer error bars include the
systematic uncertainty, added in quadrature.
In (a), (b) and (c) the PYTHIA prediction for non-singlet exchange
events is shown as open circles.
The inclusive cross section is shown in (a).  The cross section for
gap events is shown in (b) and the gap-fraction is shown in (c).
The gap-fraction is redisplayed in (d) and compared with the result of
a fit to an exponential plus a constant.
In (c) and (d) the points are
drawn at the mean \deta\ of the inclusive distribution in the
corresponding bin.
}
\end{figure}

\end{document}